\documentstyle[12pt,amsmath,amssymb,epsf,epic,cite]{article}
\numberwithin{equation}{section}

\newcommand{\be}{\begin{equation}}
\newcommand{\ee}{\end{equation}}
\newcommand{\bea}{\begin{eqnarray}}
\newcommand{\eea}{\end{eqnarray}}

\newcommand{\e}{{\rm e}}

\renewcommand{\d}{{\rm d}}

\newcommand{\grintl}{[\kern-.18em [}
\newcommand{\grintr}{]\kern-.18em ]}

\newcounter{resultcounter}[section]

\newtheorem{prop}[resultcounter]{Proposition}

\newtheorem{definition}[resultcounter]{Definition}

\def\bed{\begin{definition}}
\def\eed{\end{definition}}
\def\one{{\mathchoice {\rm 1\mskip-4mu l} {\rm 1\mskip-4mu l} {\rm 1\mskip-4.5mu l} {\rm 1\mskip-5mu l}}}

\newcommand{\s}{{\rm S}}

\newcommand{\cx}{{\mathbb C}}
\newcommand{\rx}{{\mathbb R}}
\renewcommand{\i}{{\rm i}}

\newcommand{\scalprod}[2]{\left\langle {#1}, {#2}\right\rangle}
\newcommand{\bbbone}{\mathchoice {\rm 1\mskip-4mu l} {\rm 1\mskip-4mu l}
{\rm 1\mskip-4.5mu l} {\rm 1\mskip-5mu l}}

\begin{document}

\title{Multiscale dynamics of open three-level quantum systems with two quasi-degenerate levels}

\author{Marco Merkli\footnote{Department of Mathematics and Statistics, Memorial University of Newfoundland, St. John's, NL, Canada A1C 5S7; merkli@mun.ca}\and Haifeng Song\footnote{Department of Mathematics and Statistics, Memorial University of Newfoundland, St. John's, NL, Canada A1C 5S7. Current address: Department of Mathematics, Tianjin University of Technology, Tianjin 300384, China; song\underline{\ }haifeng@126.com} \and Gennady P. Berman\footnote{Theoretical Division, Los Alamos National Laboratory, and the New Mexico Consortium, 100 Entrada Dr., Los Alamos, NM, 87544, USA; bgennady@newmexicoconsortium.org}}

\maketitle

\begin{abstract}
We consider a three-level quantum system interacting with a bosonic thermal reservoir. 
Two energy levels of the system are nearly degenerate but well separated from the third one. The system-reservoir interaction constant is larger than the energy difference of the degenerate levels, but it is smaller than the separation between the latter and the remaining level. We show that the quasi-degeneracy of energy levels leads to the existence of a manifold of quasi-stationary states, and the dynamics exhibits two characteristic time scales. On the first, shorter one, initial states approach the quasi-stationary manifold. Then, on the much longer second time scale, the final unique equilibrium is reached. 
\end{abstract}

\section{Introduction and main results}

We consider a quantum system with three energy levels interacting with a bosonic heat reservoir. One level, $E_0$, is well separated from the other two, $E\pm\sigma/2$, which are nearly degenerate. Namely, the energy gap
$$
\Delta=E_0-E>0
$$
and the level splitting $\sigma\geq 0$ satisfy $\sigma <\!\!< \Delta$. The three-level system is coupled to an (infinitely extended) bosonic heat reservoir, at temperature $T=1/\beta>0$. The system-bath interaction induces transitions between the level $E_0$ and the two almost-degenerate levels. It carries a (small) coupling constant $\lambda$. The full Hamiltonian is given by
\begin{equation}
\label{1}
H(\sigma,\lambda) = H_S(\sigma) +H_R +\lambda G\otimes\varphi(g),
\end{equation}
where
\begin{equation}
\label{2}
 H_S(\sigma) =
\left(
\begin{array}{ccc}
E_0 & 0 & 0\\
0   & E+\sigma/2 & 0\\
0   & 0 & E-\sigma/2
\end{array}
\right),
\end{equation}
\begin{equation}
\label{3}
 H_R =\sum_k \omega_k a^\dagger_k a_k
\end{equation}
and
\begin{equation}
\label{4}
 G =
\left(
\begin{array}{ccc}
0& 1 & 1 \\
1   & 0& 0\\
1   & 0 & 0
\end{array}
\right),\qquad \varphi(g)=\sum_kg_k a^\dagger_k +{\rm h.c.}
\end{equation}
The interaction is symmetric under permutation of the second and third levels, namely, the first level interacts with both of them in the same way. This symmetry facilitates the analysis, but is not required for it, see the remark after \eqref{17}.

The three-level systems with quasi-degenerate two levels (so-called $\Lambda$-systems) naturally appear in quantum optics when considering neutral atoms moving in laser fields \cite{Dil} and in bio-systems which describe the donor-acceptor exciton and electron transfer \cite{MBSa}. Recently the $\Lambda$-system was also implemented in superconducting nanocircuits \cite{Falci}.  In all these situations, it is crucial to understand the influence of the thermal bath on the quantum dynamics of the reduced density matrix. In particular, as we demonstrate in this paper, there are {\em two characteristic time-scales} (not one) describing the approach of the system to equilibrium.

We use the diagonal representation of our three-level system, denoting the orthonormal basis diagonalizing $H_S(\sigma)$ by $\{\varphi_1,\varphi_2,\varphi_3\}$.
We are interested in the regime where the system-reservoir interaction is much smaller than the gap $\Delta$ but much larger than the splitting $\sigma$,
\begin{equation}
\label{5}
0\le \sigma <\!\!< \lambda^2 <\!\!< \Delta.
\end{equation}
We call the system with $\sigma=0$ {\em degenerate}. The initial states are of the form
\begin{equation}
\label{6}
\omega_{SR} = \omega_S\otimes\omega_{R,\beta},
\end{equation}
where $\omega_S(A)={\rm Tr}(\rho_S A)$ for any observable $A$ of the 3-level system, and where  $\rho_S$ is an arbitrary initial 3-level density matrix. The bath initial state is the thermal equilibrium of the free bose gas in the thermodynamic limit (infinite volume). In this limit, the momenta of single bosons take continuous values $k\in\rx^3$. The bosonic creation and annihilation operators $a^\dagger_k$, $a_k$ (see \eqref{3}) become the infinite-volume versions $a^*(k)$, $a(k)$, satisfying the usual commutation relation $[a(k),a^*(l)]=\delta(k-l)$. The infinite-volume equilibrium is the quasi-free state determined by its two-point function
$$
\omega_{R,\beta}(a^*(k) a(l)) = \frac{\delta(k-l)}{\e^{\beta|k|}-1},
$$
in accordance with Planck's law of black body radiation. This state is obtained as the thermodynamic limit of the state given by the density matrix $\rho_R\propto e^{-\beta H_R}$. We mention that $\rho_R$ does not have a thermodynamic limit itself (as $H_R$ has continuous spectrum in that limit), but given any (quasi-local) observable $A$ of the bosons, the average ${\rm Tr}(\rho_RA)$ {\em does} have a thermodynamic limit, which equals $\omega_{R,\beta}(A)$. The state $\omega_{R,\beta}$ can be identified with a density matrix, however, the latter acts on a  {\em different} Hilbert space than the original one (``KMS construction'', see \cite{BR}). This construction is implicit in our analysis, namely, we start off with an infinitely extended reservoir.

The resonance approach we follow (also known as complex spectral deformation, complex scaling) requires certain regularity of the form factor $g(k)$, $k\in\rx^3$ (see \eqref{4}). The precise (and somewhat technical) condition on $g$ can be found in \cite{MSB}. Here, we limit ourselves to give an example of a family of form factors which are admissible: Expressed in spherical coordinates $(r,\Sigma)\in\rx^3$, $g(r,\Sigma)=r^p\e^{-r^m}g_1(\Sigma)$, where $p=-1/2+n$, $n=0,1,\ldots$, $m=1,2$ and $g_1(\Sigma)=e^{\i\phi}\bar g_1(\Sigma)$, where $\phi$ is an arbitrary phase.

The {\em reduced dynamics} of the 3-level system is obtained by tracing out the degrees of freedom of the reservoir,
\begin{equation}
\label{7.1}
\rho(t) = {\rm Tr}_{R}\big(\e^{-\i tH(\sigma,\lambda)} \rho(0)\otimes\rho_R\,\e^{\i tH(\sigma,\lambda)}\big).
\end{equation}
In this notation, it is automatically assumed that the thermodynamic limit is performed on the right side. It has been well known for a long time that the reduced dynamics is difficult to analyze \cite{BP}. A mathematically rigorous way, based on quantum resonance theory, has been developed in \cite{MSB}. It uses the general framework established in \cite{JP,BFS}. We are following this approach. In previous work the situation was considered where the system-reservoir coupling strength $\lambda$ is much smaller than all energy differences of the uncoupled small system. Except in \cite{MS}, where the opposite case is analyzed, namely $\lambda$ much larger than all system energy differences. In the present work, we consider the regime  \eqref{5} combining the two previous ones.

We are not aware of other literature on open quantum systems describing multiple time-scales due to quasi-degeneracy of energy levels. Even so, the appearence of different time-scales in open systems in different settings has been observed before. The paper \cite{Davies} examines the dynamics of a particle moving in widely separated potential wells and interacting with an infinite reservoir. The spacing of the wells and the reservoir coupling strength are related in a suitable way and two regimes appear, in which the weak coupling dynamics of the particle shows different decay characteristics and in which the set of invariant states are different.
In \cite{Alicki}, various master equations for the dynamics of a nonlinear oscillator interacting with a
reservoir are investigated. It is found (heuristically) that different generators yield more accurate descriptions
of the reduced oscillator dynamics for different time-scales. In particular, different
generators should be used for times shorter than, and longer than, the inverse of the system level spacing. In \cite{McC}, two spatially separated qubits interacting with a common thermal bath are considered. It is shown that in certain parameter regimes, the lifetime of entanglement in the qubits, created by the common bath, can be much larger than the single-qubit decoherence time.

\medskip

We now summarize our results.

\subsection{Approximate dynamics}

The resonance method yields an {\em approximation} $T_t(\rho)$ of the reduced dynamics $\rho(t)$. Its strength lies in the following two facts:  Firstly, the approximation is accurate to $O(\lambda^2)$,  {\em uniformly in time $t\geq0$}; this means that it differs from the true dynamics  by a `remainder term' which depends on time, but which has an upper bound  $O(\lambda^2)$ that is independent of  $t \ge 0$ (in particular, it does not grow as $t\rightarrow\infty$). Secondly, the derivation of the approximate dynamics and the estimates on the remainder (difference to the true dynamics) are {\em mathematically rigorous}. More precisely, we show in Section \ref{redsect} that there is a $\lambda_0>0$ s.t. if $|\lambda|<\lambda_0$, then
\begin{equation}
\label{adstat}
\sup_{t\geq 0}\| \rho(t)-T_t(\rho)\| \leq C\lambda^2,
\end{equation}
where $C$ is independent of $t$, $\lambda$ and $\sigma$. The approximate dynamics $T_t(\rho)$ satisfies the group property $T_{t+s}=T_t\circ T_s$.

As \eqref{adstat} asserts, the approximate dynamics $T_t(\rho)$ equals the true density matrix $\rho(t)$ {\em only up to an error $O(\lambda^2)$.} In particular, since $\sigma<\!\!<\lambda^2$, the approximate dynamics does not ``resolve'' $O(\sigma)$ effects in the density matrix at a fixed moment in time. However, over longer time periods, the effects on the system caused by the level splitting $\sigma$ accumulate. They eventually become larger than $O(\lambda^2)$ and our theory is able to resolve these effects. We find the time-scales over which significant (measurable) changes, larger than $O(\lambda^2)$, happen. Then, by observing these measurable changes, one can reconstruct the value of $\sigma$.  We give concrete examples of this procedure in the illustrations in section \ref{illsect} below.


\medskip

As we show, the reduced dynamics $T_t$ has an explicit form in terms of resonance energies and resonance states. This allows us to carry out a detailed analysis of the evolution, yielding the following results.

\subsection{Properties of the degenerate system ($\sigma=0$)} 

{\em (A) Multiple stationary states.}  A system state $\rho$ is called {\em stationary} or ($T_t$-){\em invariant} if $T_t(\rho)=\rho$ for all $t\geq 0$. We show that the manifold of all $T_t$-invariant states is
$$
{\cal M}_{0} = \big\{\mu\rho_{S,\beta,0}+(1-\mu)\rho_\tau\ :\ \mu\in[0,1]\big\}.
$$
Here, $\rho_{S,\beta,0}\propto\e^{-\beta H_S(0)}$ is the equilibrium state of the system with $\sigma=0$ and $\rho_\tau=|\tau\rangle\langle\tau|$, where $\tau=2^{-1/2}(\varphi_2-\varphi_3)$. Expressed in the energy basis $\{\varphi_j\}$ of $H_S$,
\begin{equation}
 \label{21'}
\rho\in{\cal M}_0\ \ \Leftrightarrow\ \
\rho=
\left(
\begin{array}{ccc}
p & 0 & 0\\
0 & \textstyle\frac12(1-p) & \alpha\\
0 & \alpha & \textstyle\frac12(1-p)
\end{array}
\right),
\end{equation}
where $p\in[0,p_{\rm max}]$ with $p_{\rm max} = (\e^{\beta\Delta}+1)^{-1}$ and $\alpha=\frac12((2\e^{\beta\Delta}+1)p-1)$.

\medskip

{\em (B) Final state dependence on initial state.} Since ${\cal M}_0$ is not a single point, the final state of the 3-level system, as $t\rightarrow\infty$, will depend on the initial condition. We show that, for any 3-level system density matrix $\rho$,
\begin{equation}
\label{20'}
\lim_{t\rightarrow\infty} T_t(\rho)= \rho_\infty =
\left(
\begin{array}{ccc}
p_\infty & 0 & 0\\
0 & \textstyle\frac12(1-p_\infty) & \alpha_\infty\\
0 & \alpha_\infty & \textstyle\frac12(1-p_\infty)
\end{array}
\right),
\end{equation}
where
\begin{equation}
p_\infty = \frac{1+[\rho ]_{11} +2{\rm Re}[\rho ]_{23}}{2(\e^{\beta\Delta}+1)} \qquad\mbox{and}\qquad
\alpha_\infty =\textstyle\frac12((2\e^{\beta\Delta}+1)p_\infty-1).
\end{equation}
The density matrix is represented here in the energy basis $\{\varphi_j\}$ of $H_S$. Of course, $\rho_\infty\in{\cal M}_0$. The convergence speed in \eqref{20'} is exponential,
\begin{equation}
\label{30'}
\| T_t(\rho)- \rho_\infty\|\leq Ce^{-\gamma_{\rm deg}\lambda^2 t},
\end{equation}
where $C$ and $\gamma_{\rm deg}>0$ (``degenerate'') are constants independent of $\lambda$ and $\|\cdot\|$ is the trace norm. The convergence rate $\gamma_{\rm deg}$ depends on the spectral density of the reservoir,
\begin{equation}
\label{specdensity}
J(\omega)=\frac12{\pi}\omega^2\int_{S^2}|g(\omega,\Sigma)|^2d\Sigma,\qquad \omega\ge0,
\end{equation}
via the quantity $J(\Delta)$ and
\begin{equation}
\label{xizero}
\widetilde J(0)  = \lim_{\omega\rightarrow 0_+}\frac{J(\omega)}{\omega}.
\end{equation}
See also point 3. in the discussion after Proposition \ref{prop1}. Let $T=1/\beta$ be the temperature. We obtain 
\begin{equation}
\label{gammadeg}
\gamma_{\rm deg}= 
\left\{
\begin{array}{ll}
\displaystyle 2\min\big\{J(\Delta)\e^{-\Delta/T}, T \widetilde J(0)\big\} \sim 2 J(\Delta)\e^{-\Delta/T} \qquad & \mbox{for $\Delta >\!\!> T$}\\
\ &\ \\
\displaystyle 2T\,\min\big\{ \frac{J(\Delta)}{\Delta}, \widetilde J(0)\big\}&  \mbox{for $\Delta <\!\!<T$}
\end{array}
\right.
\end{equation}

\subsection{Properties of the non-degenerate system ($\sigma>0$)}

{\em (A) Unique invariant state.} We show that for $\sigma>0$, the manifold of invariant states is a single point
$$
{\cal M}_{\sigma} =\{\rho_{S,\beta,\sigma}\},
$$
the equilibrium state $\rho_{S,\beta,\sigma}\propto\e^{-\beta H_S(\sigma)}$. Moreover, for every initial state $\rho$,
\begin{equation}
\label{31'}
\|T_t(\rho)-\rho_{S,\beta,\sigma}\| \leq C\e^{-\gamma_{\rm nd} t\sigma^2/\lambda^2},
\end{equation}
where $C$, $\gamma_{\rm nd}>0$ (``non-degenerate'') are constants independent of $\sigma$ and $\lambda$. Again, the convergence speed $\gamma$ depends on $J(\Delta)$. See point 3. in the discussion after Proposition \ref{prop1}. We obtain the following expressions, where $T=1/\beta$ is the temperature
\begin{equation}
\label{gammand}
\gamma_{\rm nd}= 
\left\{
\begin{array}{ll}
\displaystyle \frac12 \frac{J(\Delta)}{\vartheta^2\,\e^{\Delta/T} +J(\Delta)^2\,\e^{-\Delta/T}} &  \mbox{for $\Delta >\!\!> T$,}\\
\ &\ \\
\displaystyle   \frac38\frac{\Delta}{T}\, \frac{J(\Delta)}{\vartheta^2\Delta^2/T^2+ J(\Delta)^2}\quad &  \mbox{for $\Delta<\!\!< T$.}\\
\end{array}
\right.
\end{equation}
Here, 
$$
\vartheta =\frac12\int_{{\mathbb
R}^3}\frac{1}{1-\e^{-|k|/T}} \frac{|g(k)|^2}{|k|+\Delta}d^3k-{\rm P.V.}
\frac12\int_{{\mathbb
R}^3}\frac{1}{\e^{|k|/T}-1}\frac{|g(k)|^2}{|k|-\Delta}d^3k.
$$

The convergence to the final state, which does not depend on the initial one, takes place at a speed $\propto \sigma^2/\lambda^2$. The smaller the level spacing $\sigma$, the slower the convergence. (And indeed, for $\sigma=0$ the state $\rho_{S,\beta,\sigma}$ is not approached at all, for general initial states.)

Comparing the rates $\gamma_{\rm deg}$ and $\gamma_{\rm nd}$ we see that at low temperatures, both are exponentially small,  proportional to $\e^{-\Delta/T}$. At high temperatures, $\gamma_{\rm deg}$ grows linearly in $T$, while $\gamma_{\rm nd}$ decreases as $1/T$.

Both $\gamma_{\rm deg}$ and $\gamma_{\rm nd}$ depend on the {\em reservoir correlation time}. For example, for the form factor (see \eqref{4} and before \eqref{7.1}) 
\begin{equation}
\label{formfactor}
g(k) = A|k|^{-1/2} \e^{-\frac12 |k|/\varkappa_0}
\end{equation}
the reservoir correlation turns out to be $\tau_{\rm c}=1/\varkappa_0$ and we obtain 
\begin{equation}
\label{jjwigle}
J(\Delta) = 2\pi^2 A^2\Delta\e^{-\Delta\tau_{\rm c}}\qquad\mbox{and}\qquad \widetilde J(0) = 2\pi^2 A^2.
\end{equation}
For more details and a derivation of \eqref{jjwigle}, see Appendix \ref{appendix}. 

\medskip

{\em (B) Emergence of two time scales.} We show that the dynamics $T_t(\rho)$ has the expansion
\begin{equation}
\label{33}
T_t(\rho) = \rho_{S,\beta,\sigma} +\e^{\i t\varepsilon_0^{(2)}(\sigma,\lambda)}\chi_0^{(2)} + \sum_{\ell=1}^7 \e^{\i t\varepsilon_\ell(\sigma,\lambda)}\chi_\ell,
\end{equation}
where the $\chi$ are operators on the 3-level system (not depending on $t$) and the $\varepsilon$ are complex numbers satisfying ${\rm Im}\varepsilon\geq 0$. More precisely,
\begin{eqnarray*}
{\rm Im }\, \varepsilon_0^{(2)}  &\propto&\sigma^2/\lambda^2 +O(\lambda^2(\sigma/\lambda)^4)\\
{\rm Im }\, \varepsilon_\ell     &\propto& \lambda^2 +O((\sigma/\lambda)^2).
\end{eqnarray*}
In the considered regime $\sigma<\!\!<\lambda^2$, we have ${\rm Im }\varepsilon_0^{(2)} <\!\!< {\rm Im }\varepsilon_\ell$.  The terms in \eqref{33} proportional to $\chi_\ell$ converge to zero exponentially quickly, on a time scale $t_1\sim 1/\lambda^2$. The contribution $\chi_0^{(2)}$ survives on a much longer time scale  $t_2\sim \lambda^2/\sigma^2$. In the limit $t\rightarrow\infty$, $T_t(\rho)$ becomes the equilibrium state $\rho_{S,\beta,\sigma}$ (and \eqref{31'} follows directly from \eqref{33}).

Taking $\sigma\rightarrow 0$ in \eqref{33} gives $T_t(\rho) = \rho_{S,\beta,0} +\chi_0^{(2)} + \sum_{\ell=1}^7 \e^{\i t\varepsilon_\ell(0,\lambda)}\chi_\ell$. For $t$ larger than $\sim\lambda^{-2}$, all the exponentials in the last sum become negligible. Hence on a time scale $O(1/\lambda^2) < t < O(\lambda^2/\sigma^2)$, the state $T_t(\rho)$ is approximated by  $\rho_{S,\beta,0} +\chi_0^{(2)}$. One verifies directly (using the explicit formulas in Section \ref{lsosect})  that  $\rho_{S,\beta,0} +\chi_0^{(2)}\in{\cal M}_0$ is the point of ${\cal M}_0$ given in \eqref{21'}, associated to the initial condition at hand.

\medskip

{\bf Summary.\ } This analysis paints the following picture. For
$\sigma<\!\!<\lambda^2$, any initial system state approaches the
quasi-stationary manifold ${\cal M}_0$ on a rapid time scale
$t_1\sim1/\lambda^2$. It sojourns in a vicinity (of size
$O(\sigma)$) of the point of ${\cal M}_0$ given by \eqref{21'}
(which depends on the initial state) and finally, after
$t_2\sim\lambda^2/\sigma^2>\!\!>t_1$, it decays to the unique
equilibrium $\rho_{S,\beta,\sigma}$.

In the `usual' situation when $\lambda^2$ is smaller than all energy differences of the system alone, there is a {\em single} time scale on which the system approaches the equilibrium. This time scale is of order $1/\lambda^2$. In our situation, the almost-degeneracy of two of the system levels generates the new time scale $\sim \sigma^2/\lambda^2$. We can view this effect as a perturbation of the manifold of stationary states ${\cal M}_0$. As $\sigma>0$, this manifold is unstable and collapses to a single point $\rho_{S,\beta,\sigma}$.

\medskip

{\em Remark.\ } The rotating wave approximation and  the Born-Markov approximation cannot be used in this problem, because the energy splitting $\sigma$ is much smaller than the system-reservoir coupling constant $\lambda^2$. Our resonance approach is also not analogous to the Redfield equation. Indeed, the derivation of the Redfield equation is based on perturbation theory in $\lambda$ and as a result it can only yield characteristic decay rates $\propto \lambda^2 S(\sigma)$, where $S(\sigma)$ encodes the reservoir spectral density (and does not depend on $\lambda$). However, in the quasi-degenerate case we consider, some directions of decay have a rate $\propto \sigma^2/\lambda^2$. This functional dependence is incompatible with the Redfield approach. Rather, the emergence of rates $\propto \sigma^2/\lambda^2$ is explained as follows: our ``unperturbed system" is already interacting with the reseroir ($\lambda\neq 0$), having (complex) energies $\varepsilon\propto\lambda^2$. {\em Then} we switch on the perturbation $\sigma$ and obtain (usual second order) energy corrections $\propto \frac{\sigma^2}{\varepsilon-\varepsilon'} \propto \sigma^2/\lambda^2$.

\subsection{Illustrations} 
\label{illsect}

\begin{itemize}
\item[1.]

{\em Donor-acceptor model.\ } We view the 3-level system as a donor-acceptor system, describing, for instance, the electron transfer in chemical reactions \cite{Marcus, XuSch,MBSa}. The level $E_0$ is the donor level and $E\pm\sigma/2$ are almost-degenerate acceptor levels. Suppose the initial state is
$$
\rho_0=p_D(0)|\varphi_1\rangle\langle\varphi_1|+p_2|\varphi_2\rangle\langle\varphi_2|+p_3|\varphi_3\rangle\langle\varphi_3|,
$$
with initial populations $p_D(0)$, $p_2$ and $p_3$ of the donor and acceptor levels 2 and 3, respectively. We analyze the donor probability (c.f. \eqref{adstat} and \eqref{33})
\begin{eqnarray*}
p_D(t)&=&[\rho(t)]_{11} = [T_t(\rho_0)]_{11} +O(\lambda^2)\\
 &=& [\rho_{S,\beta,\sigma}]_{11}+\e^{\i t\varepsilon_0^{(2)}}[\chi_0^{(2)}]_{11} + \sum_{\ell=1}^7 \e^{\i t\varepsilon_\ell}[\chi_\ell]_{11}+O(\lambda^2).
\end{eqnarray*}
The initial condition is $p_D(0)\in[0,1]$. For the degenerate system, $\sigma=0$, the final state is given by \eqref{20'}, hence
$$
p_{D}(\infty) = \frac{1+p_D(0)}{2(1+\e^{\beta\Delta})}+O(\lambda^2),\qquad\mbox{for $\sigma=0$.}
$$
The final donor value depends on the initial donor value for the degenerate system.
In particular, if $p_D(0)=1$, then 
\begin{equation}
\label{donor1}
p_{D}(\infty) = (1+\e^{\beta\Delta})^{-1}+O(\lambda^2), \qquad \mbox{for $\sigma=0$.}
\end{equation}
This coincides with the value obtained in \cite{MBSa} for the multi-level acceptor model (see formula (1.15) in \cite{MBSa} with $N_A=2$, $V=0$, which gives the acceptor probability $p_A$ for a single acceptor level -- the relation to our $p_D$ being $p_D=1-2p_A$).

On the contrary, if the acceptor levels are however slightly non-degenerate (any $\sigma>0$), then the final donor probability is, independently of the initial condition, given by
\begin{eqnarray}
p_D(\infty)&=&[\rho_{S,\beta,\sigma}]_{11}+O(\lambda^2) \nonumber\\
&=& (1+\e^{\beta( \Delta+\sigma/2)} +\e^{\beta( \Delta-\sigma/2)} )^{-1}+O(\lambda^2)\nonumber\\
&=& (1+2\e^{\beta\Delta})^{-1}+O(\lambda^2),\qquad \mbox{for $\sigma>0$.}
\label{donor2}
\end{eqnarray}
To pass from the second last step to the last one, we use the fact that $\sigma<\!\!<\lambda^2$. The final donor probability is reached after a very long time $\propto \lambda^2/\sigma^2$. 

Note that the final donor probabilities \eqref{donor1} for $\sigma=0$ and \eqref{donor2} for $\sigma>0$  {\em do not coincide}.

\item[2.]
{\em  Decoherence.\ } Consider $\sigma>0$. As explained above, on the time-scale $t_1\propto 1/\lambda^2$, any initial state $\rho_0$ of the system approaches the associated quasi-stationary state 
\begin{equation}
 \label{21''}
\rho_{\rm qstat} = \left(
\begin{array}{ccc}
p & 0 & 0\\
0 & \textstyle\frac12(1-p) & \alpha\\
0 & \alpha & \textstyle\frac12(1-p)
\end{array}
\right),
\end{equation}
where 
\begin{equation}
\label{21'''}
p = \frac{1+[\rho_0]_{11} +2{\rm Re}[\rho_0]_{23}}{2(\e^{\beta\Delta}+1)} \qquad\mbox{and}\qquad
\alpha =\textstyle\frac12((2\e^{\beta\Delta}+1)p-1).
\end{equation}
More precisely, $\rho_t-\rho_{\rm qstat} =O(\lambda^2)$, for $t\approx t_1$. Thus \eqref{21''} tells us that after times $t\ge t_1$, the {\em coherences} (density matrix elements) in $\rho_t$  between the two energy subspaces of $H_S$ associated to the energies $E_0$ (non-degenerate) and $E$ (doubly degenerate), are negligibly small ($O(\lambda^2)$). However, the coherence between the two degenerate energy levels, $\alpha$, is sizeable. Even if these two levels are initially uncorrelated, $[\rho_0]_{2,3}=0$, after time $t_1$ they acquire coherence (nonzero matrix element) of the size $\alpha+O(\lambda^2)$, with $\alpha$ given in \eqref{21'''}. In particular, starting with $[\rho_0]_{2,3}=0$, we have $|[\rho_t]_{2,3}|\ge \frac14 (\e^{\beta\Delta}+1)^{-1} +O(\lambda^2)$ for  $t\sim t_1$.

The coherence between the nearly degenerate energy levels is lost (is of $O(\lambda^2)$) after time $t_2$, when the final (Gibbs) equilibrium state is reached.

\item[3.]
{\em Determining $\sigma$ from observation.\ } The basic idea is to measure the time until a property of the system changes significantly. The corresponding time scale is linked to the value of $\sigma$.  Concretely, suppose the initial 3-level system has density matrix
$$
\rho(0) = |1\rangle\langle 1| .
$$
The final state will be 
$$
\rho(\infty) = \rho_{S,\beta,\sigma} +O(\lambda^2).
$$
Consider the population of level one, 
$$
p(t)=[\rho(t)]_{1,1},\qquad \mbox{with}\quad  p(0)=1 \quad \mbox{and} \quad p(\infty)=\frac{1}{1+2\e^{\beta\Delta}} +O(\lambda^2).
$$
We use expression (1.20) to obtain
$$
{\rm Tr} \Big( \sum_{\ell=1}^7 \e^{\i t\varepsilon_\ell(\sigma,\lambda)} \chi_\ell(|\varphi_1\rangle\langle\varphi_1|)\Big) = O(\lambda^2) \mbox{\quad for times $t>\!\!>1/\lambda^2$.}
$$
This is so since all exponentials in the sum decay with rate $1/\lambda^2$. Using the formulas for $\varepsilon_0^{(2)}$ and $\chi_0^{(2)}$ given in Proposition \ref{prop1} and Section \ref{subsubresproj}, we obtain  from (1.20) that 
\begin{equation}
\label{letter1}
p(t) = \frac{1}{1+2\e^{\beta\Delta}} +\frac{\e^{-\gamma\frac{\sigma^2}{\lambda^2}t}}{2\e^{\beta\Delta}+3+\e^{-\beta\Delta}}  +O(\lambda^2),\qquad  t>\!\!>1/\lambda^2.
\end{equation}
Here, $\gamma>0$ can be read off the expression for $\varepsilon_0^{(2)}$ given in Proposition \ref{prop1}. 

\medskip

In \eqref{letter1}, the quantity decaying in time is {\em large}, $O(\lambda^0)$, for times in the window
$$
1/\lambda^2<\!\!<t <\!\!<\lambda^2/\sigma^2.
$$
Then by measuring the population $p(t)$ within this time window one obtains a measured (experimental) value for the decay rate $\tau_{\rm meas}$. We have $1/\tau_{\rm meas}=\gamma\sigma^2/\lambda^2$. 

One may also relate the value of $\lambda$ to an experimental measurement in the same way. Initially, we have $p(0)=1$. Then, observing how quickly the population decreases to its quasi-stationary value 
$$
p_{\rm qstat}= \frac{1}{1+2\e^{\beta\Delta}} +\frac{1}{2\e^{\beta\Delta}+3+\e^{-\beta\Delta}},
$$
namely the value of $p(t)$, \eqref{letter1}, where the exponential has not decayed yet significantly, yields a mesured decay time which is proportional to $1/\lambda^2$.

\item[4.]
{\em Photosynthetic Light Harvesting Complexes.\ }
Our model can be used to describe biological light harvesting complexes (LHC) of plants and algae \cite{XuSch}. In this setting, the donor is represented by a chlorophyll B molecule (ChlB), having  first excited  energy $E_0$. The acceptor is quasi-degenerate and consists of two chlorophylls A (ChlA) with slightly different excited energies $E\pm\sigma/2$, brought about by slightly different interactions with the protein environment. The donor and the two acceptors are all spatially separated from each other and so we may neglect any direct interaction. They interact only through the protein environment. The resulting coupling between donor and acceptor is relatively weak, $E-E_0>\!\!>\lambda$, while the one between the acceptor and the environment is relatively strong, $\lambda >\!\!>\sigma$. This situation often occurs in LHCs.

\end{itemize}

\section{Detailed description and proofs}

\subsection{Reduced 3-level system dynamics}
\label{redsect}
The {\em reduced density matrix} $\rho_S(t)$ of the 3-level system is defined by (compare with \eqref{7.1})
\begin{equation}
{\rm Tr}(\rho_S(t)A) = \omega_{SR}\big( \e^{\it H(\sigma,\lambda)}(A\otimes\bbbone_R)\e^{-\i tH(\sigma,\lambda)}\big),
\label{7}
\end{equation}
valid for all $A\in{\cal B}(\cx^3)$.  The starting point of the analysis is the dynamics of the uncoupled and degenerate system, $\sigma=\lambda=0$. Then the 3-level system evolves independently of the reservoir, according to the Liouville equation
$$
\textstyle\frac{d}{dt} \rho_S(t) = -i[H_S(0),\rho_S(t)].
$$
The eigenvalues of the Liouville operator $L_S(0)=[H_S(0),\,\cdot\,]$, acting on (density) matrices, are the differences of all pairs of eigenvalues of $H_S(0)$. Namely, $e_0=0$ which has degeneracy $s_0=5$ and $e_{\pm 1}=\pm\Delta$ each having degeneracy $s_{\pm 1}=2$.\footnote{For instance, the eigenvectors associated to the null space of $L_S(0)$ are $|\varphi_j\rangle\langle\varphi_j|$, $j=1,2,3$ and $|\varphi_j\rangle\langle\varphi_k|$, $(j,k)=(2,3),(3,2)$.} As the degeneracy is lifted and the interaction is switched on ($\sigma, \lambda\neq 0$) the reduced dynamics becomes complicated. It can be expressed using complex (resonance) energies $\widetilde\varepsilon_j^{(s)}(\sigma,\lambda)$, where $j=-1,0,1$ and $s=1,\ldots,s_j$. The resonance energies bifurcate out of the unperturbed values,
\begin{equation}
\label{8}
 \widetilde\varepsilon_j^{(s)}= \widetilde\varepsilon_j^{(s)}(\sigma,\lambda)= e_j +f_j^{(s)}(\sigma,\lambda),
\end{equation}
with $f_j^{(s)}(0,0)=0$. They are not analytic functions of $(\sigma,\lambda)$ at the origin of $\cx^2$. However, if one variable is held fixed away from
 zero, say $\lambda\neq 0$, then $\sigma\mapsto f_j^{(s)}(\sigma,\lambda)$ is analytic at $\sigma=0$. The same holds if the roles of $\lambda$ and
 $\sigma$ are interchanged. We have (see Sections 4 and 6 of
 \cite{MSB})
\begin{equation}
\label{9}
{\rm Tr}(\rho_S(t) A) = \sum_{j=-1,0,1}\ \sum_{s=1}^{s_j} \e^{\i t\widetilde\varepsilon_j^{(s)}} W_j^{(s)}(A) +O(\lambda^2\e^{-\alpha t}),
\end{equation}
valid for any $A\in{\cal B}(\cx^3)$ and all $t\geq 0$. Here, $W_j^{(s)}$ are linear functionals on observables $A\in{\cal B}(\cx^3)$ and the remainder term is {\em uniform} in $t\geq 0$, meaning that $|O(\lambda^2\e^{-\alpha t})|\leq C\lambda^2\e^{-\alpha t}$ for a constant $C$ independent of $\lambda, t$. We can recover the density matrix elements of $\rho_S(t)$ by choosing convenient operators $A$ in \eqref{9},
\begin{equation}
[\rho_S(t)]_{k,l} \equiv \scalprod{\varphi_k}{\rho_S(t)\varphi_l} ={\rm Tr}\big(\rho_S(t) |\varphi_l\rangle\langle\varphi_k|\big).
\label{9.1}
\end{equation}
The construction of the $\varepsilon_j^{(s)}$ and $W_j^{(s)}$ are linked to the so-called {\em level shift operators}.

\medskip

{\bf Perturbation theory and level shift operators. } It is often convenient to adopt the Liouville representation of the (system) density matrix, see for instance the book \cite{Muka}. In this representation, the density matrix $\rho_S\in{\cal B}(\cx^3)$ is given by a normalized vector $\psi_S\in\cx^3\otimes\cx^3$, such that
$$
{\rm Tr}(\rho_S A)=\scalprod{\psi_S}{(A\otimes\one_S)\psi_S}.
$$
Here, $\one_S=\one_{\cx^3}$. The original system Hilbert space ($\cx^3$) is thus doubled to obtain the Liouville Hilbert space $\cx^3\otimes\cx^3$, but in the latter the system {\em density matrix} is simply given by a {\em normalized vector}. Similarly, one has a Liouville representation for the reservoir, and one can describe the system-reservoir dynamics on the total system-reservoir Liouville  space. We give more detail about this in Section \ref{lsosect}. Here we explain how the $\varepsilon^{(s)}_j$ and the $W_j^{(s)}$ are constructed. 

The system Liouville operator acts on the system Liouville  space $\cx^3\otimes\cx^3$ as $L_S(\sigma) = H_S(\sigma)\otimes\bbbone_S-\bbbone_S\otimes H_S(\sigma)$. The spectral subspaces of $L_S(0)$ have dimension five (eigenvalue zero) and dimension two (for each eigenvalue $\pm\Delta$). On these eigenspaces, labelled by $j=0,\pm 1$, act the {\em level shift operators} $\Lambda_j$, which can be written in diagonal form as
\begin{equation}
\label{10}
\Lambda_j = \sum_{s=1}^{s_j}  \varepsilon_j^{(s)} P_j^{(s)},
\end{equation}
where $\varepsilon_j^{(s)}=\varepsilon_j^{(s)}(\sigma,\lambda)$ are the eigenvalues of $\Lambda_j$, satisfying ${\rm Im}\,\varepsilon_j^{(s)}\geq 0$, and $P_j^{(s)}$ are the spectral rank-one eigenprojections, which are generally not orthogonal. The level shift operators \eqref{10} describe the corrections to the eigenvalues of $L_S(0)$ under the perturbation $\sigma$ and $\lambda$.

The eigenvalues of the level shift operators are related to $\widetilde\varepsilon_j^{(s)}$, \eqref{8}, by
\begin{equation}
\label{11}
\widetilde\varepsilon_j^{(s)} = \varepsilon_j^{(s)}+ O(\lambda^2(|\lambda|+\sigma)).
\end{equation}
The functionals $W_j^{(s)}$ are related to the eigenprojections by
\begin{eqnarray}
W_j^{(s)} (A) &=& \chi_j^{(s)}(A) +O(\lambda^2),\nonumber\\
\widetilde\chi_j^{(s)}(A) &=& \scalprod{\psi_S}{ B P_j^{(s)} (A\otimes\one_S)\psi_{\rm ref}}.
\label{12}
\end{eqnarray}
The functionals $\widetilde\chi_j^{(s)}$ equal, up to a small modification explained in Section \ref{approxdyn}, the $\chi_j^{(s)}$ in the expansion \eqref{33}. Note that $\widetilde\chi_j^{(s)}$ depends on the initial state.
In the expression on the right side of \eqref{12}, we have used the Liouville representation of the 3-level system. The {\em reference vector} $\psi_{\rm ref}$ in \eqref{12} is the {\em trace state} of the 3-level system, given by
\begin{equation}
\label{refstate}
\psi_{\rm ref} = \textstyle\frac{1}{\sqrt 3}\displaystyle \sum_{j=1}^3 \varphi_j\otimes\varphi_j.
\end{equation}
The operator $B$ in \eqref{12} is of the form $B=\one_\s\otimes b$, with $b\in{\mathcal B}(\cx^3)$, satisfying
\begin{equation}
\label{Bop}
B\psi_{\rm ref} = \psi_S.
\end{equation}
Given any $\psi_S$, such a $B$ exists and is unique -- this is known as a property of `cyclicity and separability' of the reference state
$\psi_{\rm ref}$, see e.g. \cite{BR}. For instance, if the initial state of the 3-level system is $\rho_S=|\varphi_1\rangle\langle\varphi_1|$, then
$\psi_S=\varphi_1\otimes\varphi_1$ and $b=\sqrt{3}|\varphi_1\rangle\langle\varphi_1|$.

\subsubsection{The approximate dynamics}

\label{approxdyn}

Combining   \eqref{9}, \eqref{11} and \eqref{12} we obtain
\begin{equation}
\label{15}
\rho_S(t) = \widetilde T_t(\rho_S) +O(\lambda^2),
\end{equation}
where the remainder is independent of $t\geq 0$ and $\sigma$, and where  $\widetilde T_t$ is defined by the relation
\begin{equation}
\label{14}
{\rm Tr}(\widetilde T_t(\rho) A) := \sum_{j=-1,0,1}\sum_{s=1}^{s_j} \e^{\i t\varepsilon_j^{(s)}} \widetilde \chi_j^{(s)}(A),
\end{equation}
valid for all 3-level system observables $A$ and all 3-level system density matrices $\rho$. We now refine this dynamics to obtain the correct final state. As one checks directly (see e.g. Section \ref{subsubresproj}),
$$
\widetilde\chi_0^{(1)}(A) = {\rm Tr}(\rho_{S,\beta,0}A)
$$
is the system equilibrium state with $\sigma=0$. The limit $t\rightarrow\infty$ of $\widetilde T_t(\rho)$, \eqref{14}, is precisely $\widetilde\chi_0^{(1)}(A)$ (for $\sigma>0$, all exponentials but one have strictly positive imaginary part, see Proposition \ref{prop1} below -- note also that this means that $\rho_{S,\beta,0}$ is a stationary state, a fact one can establish directly using the resonance data, see Subsections \ref{subsupproofprop2} and \ref{subsubstat}). On the other hand, the true asymptotic state, as $t\rightarrow\infty$ of the system is the reduction of the coupled system-reservoir equilibrium state, reduced to $S$ alone (partial trace). That state is $\rho_{S,\beta,\sigma}+O(\lambda^2)$.  Therefore, we will now improve the approximate dynamics $\widetilde T_t$ to a slightly modified one, $T_t$, having the property that its final state is $\rho_{S,\beta,\sigma}$ for $\sigma\geq0$, not $\rho_{S,\beta,0}$. To do so, we note that
$$
\rho_{S,\beta,\sigma} = X_\sigma \rho_{S,\beta,0} X_\sigma,
$$
where
$$
X_\sigma =\sqrt{\frac{Z_{S,\beta,0}}{Z_{S,\beta,\sigma}}}\big( |\varphi_1\rangle\langle\varphi_1| + \e^{-\beta\sigma/4}|\varphi_2\rangle\langle\varphi_2| + \e^{\beta\sigma/4}|\varphi_3\rangle\langle\varphi_3|\big),
$$
with $Z_{S,\beta,\sigma}={\rm Tr}\rho_{S,\beta,\sigma}$ the system partition function. We now set
\begin{equation}
\label{thechi}
\chi_j^{(s)}(A) := \widetilde \chi_j^{(s)}(X_\sigma AX_\sigma)
\end{equation}
and define the improved approximate dynamics $T_t$ by
\begin{equation}
\label{thenewdyn}
{\rm Tr}(T_t(\rho)A) = \sum_{j=-1,0,1}\sum_{s=1}^{s_j}\e^{\i t\varepsilon_j^{(s)}}\chi_j^{(s)}(A).
\end{equation}
Since $X_\sigma=\bbbone +O(\sigma)$ and $\sigma<\!\!<\lambda^2$, we have from \eqref{15} and \eqref{thechi} that still
\begin{equation}
\label{new15}
\rho_S(t) = T_t(\rho_S) +O(\lambda^2),
\end{equation}
uniformly in $t\ge 0$. This is the statement \eqref{adstat}. The final state ($\sigma>0$) of $T_t$ is
\begin{equation}
\label{thenewfinstate}
\chi_0^{(1)}(A) = \widetilde\chi_0^{(1)}(X_\sigma AX_\sigma)={\rm Tr}(\rho_{S,\beta,0}X_\sigma AX_\sigma) = {\rm Tr}(\rho_{S,\beta,\sigma}A).
\end{equation}

Setting
\begin{equation}
\label{24}
U(t)=\sum_{j=-1,0,1}\sum_{s=1}^{s_j}\e^{\i t\varepsilon_j^{(s)}} P_j^{(s)},
\end{equation}
where the $P_j^{(s)}$ are the spectral projections of the level shift operator, \eqref{10}, and taking into account \eqref{12} and \eqref{thechi}, we get
\begin{equation}
\label{23}
{\rm Tr}\big(T_t(\rho) A\big)= \scalprod{B^*\psi_S}{U(t)(X_\sigma AX_\sigma\otimes\bbbone_S)\psi_{\rm ref}}.
\end{equation}
Since $\psi_{\rm ref}$ is cyclic, we have that for any $t\geq 0$ and any operator $A$, there exists a unique operator $\alpha_t(A)$ satisfying
\begin{equation}
\label{25}
U(t)(A\otimes\bbbone_S)\psi_{\rm ref} = (\alpha_t(A)\otimes\bbbone_S)\psi_{\rm ref}.
\end{equation}
This defines the {\em Heisenberg evolution} $t\mapsto \alpha_t(A)$. Since the $P_j^{(s)}$ are spectral projections, they satisfy $P_j^{(s)}P_{j'}^{(s')}=\delta_{j,j'}\delta_{s,s'}P_j^{(s)}$, from which it follows that $U(t+s)=U(t)U(s)$, which implies that $\alpha_{t+s}(A)=\alpha_t(\alpha_s(A))$. It follows from \eqref{23}, \eqref{25} and the fact that $B=\bbbone_S\otimes b$ commutes with $\alpha_t(A)\otimes\bbbone_S$ and $B\psi_{\rm ref}=\psi_0$ that
\begin{equation}
\label{26}
{\rm Tr}\big(T_t(\rho)A\big) = {\rm Tr}\big(\rho \alpha_t(X_\sigma AX_\sigma)\big).
\end{equation}
We derive from \eqref{26} that the map $\rho\mapsto T_t(\rho)$ is linear: for all $t\geq 0$, $A\in{\cal B}({\cx^3})$, all density matrices $\rho$, $\rho'$ and all $z,z'\in\cx$,
\begin{eqnarray}
{\rm Tr}\big( T_t(z\rho+z'\rho')A\big) &=& {\rm Tr}\big( (z\rho+z'\rho') \alpha_t(X_\sigma AX_\sigma)\big)\nonumber\\
&=& z {\rm Tr}\big(\rho\alpha_t(X_\sigma AX_\sigma )\big) +z' {\rm Tr}\big(\rho'\alpha_t(X_\sigma AX_\sigma )\big)\nonumber\\
&=& z {\rm Tr}\big(T_t(\rho) A\big) +z'{\rm Tr}\big(T_t(\rho') A\big)\nonumber\\
&=& {\rm Tr}\big( (zT_t(\rho)+z'T_t(\rho')) A\big).
\label{lin}
\end{eqnarray}

The following result analyzes the resonance energies
$\varepsilon_j^{(s)}$ appearing in $U(t)$. Recall that the square
integrable function $g(k)=g(\omega,\Sigma)$ (spherical coordinates)
is the form factor \eqref{1}. Recall the definition of the spectral density, \eqref{specdensity} and \eqref{xizero}. Define
\begin{eqnarray}
\delta &=& 2\frac{J(\Delta)}{\e^{\beta\Delta}-1}\ge 0\label{delta}\\
\vartheta &=&\frac12\int_{{\mathbb
R}^3}(1+\mu_\beta(k))\frac{|g(k)|^2}{|k|+\Delta}d^3k-{\rm P.V.}
\frac12\int_{{\mathbb
R}^3}\mu_\beta(k)\frac{|g(k)|^2}{|k|-\Delta}d^3k\in {\mathbb R}\label{theta}
\end{eqnarray}
where $\mu_\beta(k)=(\e^{\beta|k|}-1)^{-1}$.

\begin{prop}[Resonance energies]
\label{prop1}
Assume that $0<\sigma <\!\!<\lambda^2 <\!\!<\Delta$. The resonances bifurcating out of the origin are given by
\begin{eqnarray*}
\varepsilon_0^{(1)} &=& 0\\
\varepsilon_0^{(2)} &=& \i\delta\frac{(2+\e^{-\beta\Delta})}{2(1+\e^{-\beta\Delta})(4 \vartheta^2+\delta^2)}\frac{\sigma^2}{\lambda^2} +O(\lambda^2(\sigma/\lambda)^4)\\
\varepsilon_0^{(3)} &=& 2\i\delta \lambda^2 (1+\e^{\beta \Delta}) +O((\sigma/\lambda)^2)\\
\varepsilon_0^{(4)} &=& \i\delta\lambda^2 +2\lambda^2 \vartheta+O((\sigma/\lambda)^2)\\
\varepsilon_0^{(5)} &=& \i\delta\lambda^2 -2\lambda^2\vartheta+O((\sigma/\lambda)^2).
\end{eqnarray*}
The resonances bifurcating out of the unperturbed energy $\Delta$ are
\begin{eqnarray*}
\varepsilon_1^{(1)} &=& \Delta -\lambda^2 {\rm P.V.}\int_{{\mathbb
R}^3}\frac{|g(k)|^2}{|k|}d^3k+2\i\lambda^2\frac{\widetilde J(0)}{\beta} +O((\sigma/\lambda)^2)\\
\varepsilon_1^{(2)}
&=& \Delta+4\i\lambda^2\frac{\widetilde J(0)}{\beta}+ O((\sigma/\lambda)^2).
\end{eqnarray*}
Finally, the resonances $\varepsilon_{-1}^{(1)}$ and $\varepsilon_{-1}^{(2)}$, bifurcating out of $-\Delta$, are obtained from the expressions for $\varepsilon_1^{(1)}$ and $\varepsilon_1^{(2)}$ by replacing $\Delta$ with $-\Delta$.
\end{prop}

{\bf Discussion.\ } 1. More precisely, by $0<\sigma <\!\!<\lambda^2 <\!\!<\Delta$ in the lemma, we mean the following. There is a constant $\lambda_0$ (much smaller than $\sqrt\Delta$), such that for any $\lambda$ with $|\lambda|<\lambda_0$ there is another constant $\sigma_0$ (depending on $\lambda$), such that the result holds for all $0<\sigma<\sigma_0$.

2. One resonance, $\varepsilon_0^{(1)}$, is always zero and $\varepsilon_0^{(2)}$ is very close to the origin ($\sigma$ small). The other three resonances bifurcating out of the origin, $\varepsilon_0^{(j)}$, $j=3,4,5$, are at a distance $O(\lambda^2)$ from the origin. The resonances bifurcating out of $\pm\Delta$ have an imaginary part $O(\lambda^2)$ in the upper half plane.

3. The convergence speeds $\gamma_{\rm deg}$ and $\gamma_{\rm nd}$ in formulas \eqref{30'} and \eqref{31'} are obtained as follows. For $\sigma=0$ we have 
$$
\gamma_{\rm deg}=\min\{{\rm Im}\varepsilon_j^{(s)}, (j,s)\neq (0,1),(0,2)\} = \min\{J(\Delta)/(\e^{\beta\Delta}-1),\widetilde J(0)/\beta\}.
$$
For $\sigma>0$ we have $\gamma_{\rm nd}={\rm Im}\varepsilon_0^{(2)}$.

\subsection{The degenerate system, $\sigma=0$}

A state $\omega$ of the 3-level system and reservoir together is said to be stationary (for the degenerate Hamiltonian $H=H(\sigma=0,\lambda)$) if $\omega(\e^{\i tH} A\e^{-\i tH})=\omega(A)$ for all times $t$ and all system-reservoir observables $A$. The degenerate system has two stationary states. One is the {\em joint} equilibrium state, in which the 3-level system in entangled with the reservoir due to the interaction. The other stationary state is the product state
\begin{equation}
\label{16}
\omega_{SR,0}=\scalprod{\tau}{\cdot\,\tau}\otimes\omega_{R,\beta},
\end{equation}
where $\omega_{R,\beta}$ is the reservoir equilibrium and
\begin{equation}
\label{17}
\tau =2^{-1/2}(\varphi_2-\varphi_3).
\end{equation}
It is readily seen that $H_S(0)\tau=E\tau$ and $G\tau=0$, from which stationarity of $\omega_{SR,0}$ follows.

{\em Remarks.\ } 1. We note that even if $G$ was not symmetric in the levels two and three, i.e., if $G$ was of the form
\begin{equation}
\label{Ge}
G_\gamma =
\left(
\begin{array}{ccc}
0 & 1 & \gamma\\
1 & 0 & 0\\
\bar\gamma & 0 & 0
\end{array}
\right)
\end{equation}
for some $\gamma\in\cx$, $\gamma\neq 1$, the degenerate system would still have two stationary states. One is again the coupled equilibrium and the other one is of the form \eqref{16} with $\tau$ replaced by $\tau_\gamma\propto \gamma\varphi_2-\varphi_3$.

2. One may consider models where the two (quasi-)degenerate levels are coupled by a matrix element $v$ in the coupling operator $G$ or $G_\gamma$. (This means, in the matrix \eqref{Ge}, replace the zeroes by $v$ and $\bar v$ in the $(2,3)$ and $(3,2)$ entries, respectively.) Equivalently, one could introduce a direct matrix element $\widetilde v$ between the two (quasi-)degenerate levels in the Hamiltonian \eqref{1}. In the regime $\sigma,\widetilde v<\!\!<\lambda^2<\!\!<\Delta$, the system dynamics still exhibits two time scales, because there are still two almost degenerate levels, lying far from the third one (relative to the size of $\lambda$). For $\widetilde v>\!\!>\lambda^2$, all three system energy levels are well separated and we are in the usual regime where only one decay time scale ($\propto 1/\lambda^2$) appears.

\medskip

As a consequence of the non-uniqueness of the stationary state, the long-time behaviour of the system depends on the initial condition.

\subsubsection{Final states and stationary states of the 3-level system}

Let $\rho$ be an (initial) state of the 3-level system. According to \eqref{thenewdyn} we have
\begin{equation}
\label{-1}
\lim_{t\rightarrow\infty} {\rm Tr}\big(T_t(\rho)A\big) = \chi_0^{(1)}(A) +\chi_0^{(2)}(A)
\end{equation}
for any observable $A$.
Indeed, all $\varepsilon_j^{(s)}$ with $j=\pm1$ and with $j=0$ and $s\geq 3$ have strinctly positive imaginary part (see Proposition \ref{prop1}),
 so that the corresponding exponentials decay at $t\rightarrow \infty$. The functionals $w_0^{(1)}$, $w_0^{(2)}$ can be expresed via the projections onto
 the resonance state associated to the resonance energies $\varepsilon_0^{(1)}=\varepsilon_0^{(2)}=0$ (see \eqref{2.36}). We obtain the following result.
\begin{prop}[Final state]
\label{prop2}
Consider $\sigma=0$ (degenerate system). Let $\rho_0$ be an arbitrary 3-level density matrix . Then
\begin{equation}
\label{20}
\lim_{t\rightarrow\infty} T_t(\rho_0)= \rho_\infty =
\left(
\begin{array}{ccc}
p_\infty & 0 & 0\\
0 & \textstyle\frac12(1-p_\infty) & \alpha_\infty \\
0 & \alpha_\infty & \textstyle\frac12(1-p_\infty)
\end{array}
\right),
\end{equation}
where
\begin{equation}
p_\infty = \frac{1+[\rho_0]_{11} +2{\rm Re}[\rho_0]_{23}}{2(\e^{\beta\Delta}+1)} \qquad\mbox{and}\qquad
\alpha_\infty =\textstyle\frac12((2\e^{\beta\Delta}+1)p_\infty-1).
\end{equation}
(Here, $[\rho_0]_{ij}$ is the matrix element of $\rho_0$, see \eqref{9.1}). The speed of convergence is  exponential,
\begin{equation}
\label{30}
\| T_t(\rho_0)- \rho_\infty\|\leq Ce^{-c\lambda^2 t},
\end{equation}
where $C$ and $c>0$ are constants independent of $\lambda$ and $\|\cdot\|$ is the trace norm.
\end{prop}

{\bf Discussion.\ } 1. The final density matrix depends on the parameter $p_\infty$, which involves the initial density matrix through the matrix elements $[\rho]_{11}$ and ${\rm Re}[\rho]_{23}$ only.

2. There are no coherences between the non-degenerate and the two degenerate levels in the final density matrix (block diagonal). The final state is symmetric with respect to interchanging levels two and three.

3. The convergence speed is determined by the imaginary part of the non-zero resonances, which is $\propto \lambda^2$.

We give a proof of Proposition \ref{prop2} in Section \ref{subsupproofprop2}

\medskip

Suppose that $\rho$ is stationary for $T_t$, i.e., that $T_t(\rho)=\rho$ for all $t\geq 0$. Then clearly $\rho$ has to be of the form $\rho_\infty$ as in \eqref{20}. The converse is true too, as shows the next result.

\begin{prop}[Invariant states]
\label{prop3}
Consider $\sigma=0$ (degenerate system).  A density matrix $\rho$ of the 3-level system satisfies $T_t(\rho)=\rho$ for all $t\geq 0$ if and only if it has the form
\begin{equation}
\label{21}
\rho=
\left(
\begin{array}{ccc}
p & 0 & 0\\
0 & \textstyle\frac12(1-p) & \alpha\\
0 & \alpha & \textstyle\frac12(1-p)
\end{array}
\right),
\end{equation}
where $p\in[0,p_{\rm max}]$ and $\alpha=\frac12((2\e^{\beta\Delta}+1)p-1)$ and $p_{\rm max} = (\e^{\beta\Delta}+1)^{-1}$.
\end{prop}

To see why Proposition \ref{prop3} holds, we first note that both density matrices
$$
\rho_{S,\beta,0}=\frac{\e^{-\beta H_S(0)}}{{\rm Tr}\e^{-\beta H_S(0)}}\qquad \mbox{and}\qquad \rho_\tau=|\tau\rangle\langle\tau|
$$
are stationary for $T_t$, namely
\begin{equation}
\label{29}
T_t(\rho_{S,\beta,0})=\rho_{S,\beta,0},\qquad T_t(\rho_\tau)=\rho_\tau.
\end{equation}
We show \eqref{29} directly in Section \ref{subsubstat}. Heuristically, stationarity of $\rho_\tau$ follows from the fact that the product state \eqref{16} is stationary for the total dynamics. The Gibbs state $\rho_{S,\beta,0}$ is stationary since it is the reduction of the full system-reservoir equilibrium state, up to corrections of $O(\lambda^2)$, and the dynamics $T_t$ is an approximation of the true 3-level system dynamics which is accurate to $O(\lambda^2)$, {\em uniformly} for all times $t\geq 0$.

Since both $\rho_{S,\beta,0}$ and $\rho_\tau$ are stationary, the family
\begin{equation}
\label{22}
\rho(\mu)=\mu\rho_{S,\beta,0} +(1-\mu)\rho_\tau, \quad \mu\in[0,1],
\end{equation}
has to be stationary as well. This is so since $\rho\mapsto T_t(\rho)$ is linear, see \eqref{lin}.

Written in the basis $\varphi_j$, $j=1,2,3$, the density matrix $\rho(\mu)$, \eqref{22}, has exactly the form \eqref{21}, with $p=\mu(1+2\e^{\beta\Delta})^{-1}$. So $\rho$, \eqref{21}, is $T_t$-invariant. This shows Proposition \ref{prop3}.

\subsection{The non-degenerate system, $\sigma>0$}

As the level-splitting is lifted, for $\sigma>0$, the state $\omega_{RS,0}$, \eqref{16}, is not stationary any more. However, the joint equilibrium is still stationary, of course.

\begin{prop}[Unique final state]
\label{prop4} Suppose the spectral density satisfies $J(\Delta)>0$ and $\widetilde J(0)>0$, see \eqref{specdensity} and \eqref{xizero}. Then, for $\sigma>0$, every initial 3-level density
matrix $\rho$ converges to the Gibbs state,
$$
\lim_{t\rightarrow\infty} T_t(\rho) = \rho_{S,\beta,\sigma}=\frac{\e^{-\beta H_S(\sigma)}}{{\rm Tr}\e^{-\beta H_S(\sigma)}}.
$$
The convergence is exponential,
\begin{equation}
\label{31}
\|T_t(\rho)-\rho_{S,\beta,\sigma}\| \leq C\e^{-ct\sigma^2/\lambda^2},
\end{equation}
where $C$, $c>0$ are constants independent of $\sigma$ and $\lambda$.
\end{prop}

{\bf Discussion.\ } 1. The positivity condition on the spectral density, $J(\Delta), \widetilde J(0)>0$, ensures that the levels are coupled effectively to the reservoir. They imply that all resonance energies (except zero) have strictly positive imaginary part, see Proposition \ref{prop1}.

2. The equilibrium is approached with a convergence speed proportional to the imaginary part of the resonance having the smallest nonzero imaginary part, $\varepsilon_0^{(2)}$, which is $\propto \sigma^2/\lambda^2$ (see Proposition \ref{prop1}). As $\sigma\rightarrow 0$, the right side of \eqref{31} does not decay to zero as $t\rightarrow\infty$. This reflects the fact that there are invariant states {\em other} than $\rho_{S,\beta,0}$ in the degenerate case.

3. The time scale $t\sim t_2=\lambda^2/\sigma^2$ (c.f. \eqref{31}) is {\em much larger} than $t\sim t_1=1/\lambda^2$ (c.f. \eqref{30}), in our parameter regime $\sigma<\!\!<\lambda^2$.

\subsection{Resonance data and proofs of Propositions \ref{prop1} and \ref{prop2}}
\label{lsosect}

\subsubsection{The Liouville space}

In the thermodynamic limit, the equilibrium state of the bosonic reservoir is represented as a vector in a new Hilbert space, called the Liouville- or GNS-Hilbert space. This setup has been discussed in \cite{MSB}. We only outline the parts that are needed for us to define the level shift operators, whose eigenvalues and eigenvectors encode the reduced dynamics as in Section \ref{redsect}.

The Hilbert space is
$$
{\cal H} = \cx^3\otimes\cx^3\otimes{\cal F},
$$
where $\cx^3\otimes\cx^3$ is the Liouville space of the 3-level system and
$$
{\cal F} = \bigoplus_{n\geq 0}L^2_{\rm symm}\big( (\rx\times S^2)^n,(\d u\times\d\Sigma)^n\big)
$$
is the reservoir space, the symmetric Fock space over the single-particle space $L^2(\rx\times S^2)$. On $\cal H$, the Liouville operator is defined as
$$
L(\sigma,\lambda)=L_S(\sigma)+L_R+\lambda I,
$$
where $L_S(\sigma)=H_S(\sigma)\otimes \bbbone_S- \bbbone_S \otimes H_S(\sigma)$ and $L_R=\d\Gamma(u)$ is the second quantization of the multiplication operator by $u$, acting on $\cal F$. The interaction is given by $I=G\otimes\bbbone_S\otimes\varphi(g_\beta)-\bbbone_S\otimes G\otimes
J_R\e^{-\beta L_R/2}\varphi(g_\beta) J_R\e^{-\beta L_R/2}$. Here, $\varphi(g_\beta)$ is the field operator acting on $\cal F$, smoothed out with the positive temperature form factor
$$
g_\beta(u,\Sigma):= \sqrt{\frac{u}{1-\e^{-\beta u}}}|u|^{1/2}\left\{
\begin{array}{ll}
g(u,\Sigma) & \mbox{if $u\geq 0$,}\\
-\overline g(-u,\Sigma) & \mbox{if $u<0$.}
\end{array}
\right.
$$
On the right side, $g\in L^2(\rx^3)$ is represented in spherical coordinates $(u,\Sigma)\in\rx_+\times S^2$. The operators $J_R$ and $\Delta_R$ in the definition of $I$ above are the modular conjugation. Its action is given by $J_R\Psi(u_1,\Sigma_1,\ldots, u_n,\Sigma_n) = \overline{\Psi(-u_1,\Sigma_1,\ldots,-u_n,\Sigma_n)}$, linearly extended to all of $\cal F$.

\subsubsection{The level shift operators}

Let $e\in\{0,\pm \Delta\}$ be an eigenvalue of $L_S(0)$. We denote by $P_e$ the projection $\chi_{L_S=e}\otimes P_\Omega$, where $P_\Omega=|\Omega\rangle\langle\Omega|$ is the projection onto the vacuum $\Omega\in \cal F$ and $\chi_{L_S=e}$ is the spectral projection of $L_S$ onto the eigenspace associated with $e$. Note that $\dim P_0=5$ while $\dim P_{\pm\Delta}=2$.  We define the level shift operators
\begin{equation}
\label{lso}
\Lambda_e =  P_e\Big( L(\sigma,\lambda) - L(\sigma,\lambda)\overline P_e (\overline{L(0,0)}-e+\i0^+)^{-1}\overline P_e L(\sigma,\lambda)\Big)P_e,
\end{equation}
where $\overline P_e =\bbbone -P_e$ and $\overline{L(0,0)} = \overline P_e L(0,0) \overline P_e\upharpoonright_{{\rm Ran}\overline P_e}$.

{\bf The level shift operator $\Lambda_0$.\ } We set
$$
\nu=\frac{1}{\sqrt 2}\left[
   \begin{array}{c}
     0 \\
     1\\
    1 \\
   \end{array}
 \right]\qquad\mbox{and}\qquad
  \tau=\frac{1}{\sqrt 2}\left[
   \begin{array}{c}
     0 \\
     1\\
    -1 \\
   \end{array}
 \right].
 $$
Then
 \begin{equation}
\label{basis}
\{\Psi_1=\varphi_{11}, \Psi_2=\nu\otimes\nu, \Psi_3=\tau\otimes\tau, \Psi_4=\frac{1}{\sqrt 2}(\varphi_{23}-\varphi_{32}), \Psi_5=\frac{1}{\sqrt 2}
 (\varphi_{22}-\varphi_{33})\}
\end{equation}
is an orthonormal basis of Ran$P_0$, and  $\Lambda_0$ is represented as a $5\times 5$ matrix in this basis. Note that $\Lambda_0=B+A$, where
 \begin{equation}
 \begin{split}
 B=&P_0L(\sigma,\lambda)P_0\\
 A=&-P_0L(\sigma,\lambda)\overline P_0(\overline{L(0,0)}+\i0^+)^{-1}\overline P_0 L(\sigma,\lambda)P_0.
 \end{split}
 \end{equation}
 The matrix $B$ is easily found to be
 \begin{equation}\label{B}
 B={\sigma}\left(
    \begin{array}{ccccc}
      0 & 0 & 0 & 0 & 0 \\
    0 & 0 &  0 & \frac{1}{\sqrt 2} & 0 \\
     0 & 0 & 0 & -\frac{1}{\sqrt 2}& 0 \\
    0 & \frac{1}{\sqrt 2} & -\frac{1}{\sqrt 2} & 0 & 0 \\
   0 & 0 & 0 & 0 & 0 \\
   \end{array}
    \right).
\end{equation}

To obtain the matrix elements of $A$, we first calculate
 \begin{equation}
 \begin{split}
 A \Psi_1=&-2\lambda^2\Psi_1\{\langle \varphi(L_R-\Delta+\i0^+)^{-1}\varphi\rangle+\langle \tilde{\varphi}(L_R+\Delta+\i0^+)^{-1}\varphi\rangle\}\\
 &+2\lambda^2\Psi_2\{\langle \varphi(L_R+\Delta+\i0^+)^{-1}\varphi\rangle+\langle \tilde{\varphi}(L_R-\Delta+\i0^+)^{-1}\varphi\rangle\},\\
 A \Psi_2=&2\lambda^2\Psi_1\{\langle \varphi(L_R-\Delta+\i0^+)^{-1}\varphi\rangle+\langle \tilde{\varphi}(L_R+\Delta+\i0^+)^{-1}\varphi\rangle\}\\
 &-2\lambda^2\Psi_2\{\langle \varphi(L_R+\Delta+\i0^+)^{-1}\varphi\rangle+\langle \tilde{\varphi}(L_R-\Delta+\i0^+)^{-1}\varphi\rangle\}\\
 =&-A\Psi_1,\\
  A\Psi_3=&0,\\
   A\Psi_4=&-\lambda^2\Psi_4\{\langle \varphi(L_R+\Delta+\i0^+)^{-1}\varphi\rangle+\langle \tilde{\varphi}(L_R-\Delta+\i0^+)^{-1}\varphi\rangle\}\\
   &+\lambda^2\Psi_5\{\langle \varphi(L_R+\Delta+\i0^+)^{-1}\varphi\rangle-\langle \tilde{\varphi}(L_R-\Delta+\i0^+)^{-1}\varphi\rangle\}\\
  A\Psi_5=&-\lambda^2\Psi_5\{\langle \varphi(L_R+\Delta+\i0^+)^{-1}\varphi\rangle+\langle \tilde{\varphi}(L_R-\Delta+\i0^+)^{-1}\varphi\rangle\}\\
   &+\lambda^2\Psi_4\{\langle \varphi(L_R+\Delta+\i0^+)^{-1}\varphi\rangle-\langle \tilde{\varphi}(L_R-\Delta+\i0^+)^{-1}\varphi\rangle\}\\
 \end{split}
 \end{equation}
Here $\langle \cdot \rangle=\langle \Omega,\cdot\, \Omega\rangle$.
Let
\begin{equation}
\begin{split}
\xi=&\langle \varphi(L_R-\Delta+\i0^+)^{-1}\varphi\rangle,\,\,\,\,\, \xi_-=\langle \varphi(L_R+\Delta+\i0^+)^{-1}\varphi\rangle,\\
\tilde\xi=&\langle \tilde{\varphi}(L_R-\Delta+\i0^+)^{-1}\varphi\rangle, \,\,\,\,\,  \tilde\xi_-=\langle \tilde{\varphi}(L_R+\Delta+\i0^+)^{-1}\varphi\rangle,\\
\end{split}
\end{equation}
then we have
\begin{equation}
 \begin{split}
  A \Psi_1=&-2\lambda^2\Psi_1(\xi+\tilde\xi_-)+2\lambda^2\Psi_2(\xi_-+\tilde\xi)\\
 A \Psi_2=&+2\lambda^2\Psi_1(\xi+\tilde\xi_-)-2\lambda^2\Psi_2(\xi_-+\tilde\xi)\\
  A \Psi_3=&0\\
 A \Psi_4=&-\lambda^2\Psi_4(\xi_-+\tilde\xi)-\lambda^2\Psi_5(\tilde\xi-\xi_-)\\
 A \Psi_5=&-\lambda^2\Psi_4(\tilde\xi-\xi_-)-\lambda^2\Psi_5(\tilde\xi+\xi_-).\\
\end{split}
\end{equation}
It follows from \eqref{delta} and \eqref{theta} that
\begin{equation}
\begin{split}
\tilde\xi+\xi_-=&-2\i\lim_{\epsilon\rightarrow 0^+}\langle \varphi\frac{\epsilon}{(L_R+\Delta)^2+\epsilon^2}\varphi\rangle=-\i\delta\\
\xi+\tilde\xi_-=&-2\i\lim_{\epsilon\rightarrow 0^+}\langle \varphi\frac{\epsilon}{(L_R-\Delta)^2+\epsilon^2}\varphi\rangle=-\i\e^{\beta\Delta}\delta\\
\xi_--\tilde\xi=&2\vartheta.
\end{split}
\end{equation}
So the matrix representation of the operator $A$  in the basis
\eqref{basis} is
\begin{equation}\label{A}
 A=
\left(
    \begin{array}{ccccc}
      \i 2\e^{\beta \Delta}\delta\lambda^2 & -\i 2\e^{\beta \Delta}\delta\lambda^2 & 0 & 0 & 0 \\
    -\i2\delta\lambda^2 & \i2\delta\lambda^2 & 0 & 0 & 0\\
     0& 0 & 0 & 0 & 0 \\
    0 & 0 & 0 & {\i\delta\lambda^2} & {2\lambda^2 \vartheta} \\
   0 & 0 & 0 & {2\lambda^2 \vartheta} & {\i\delta\lambda^2} \\
   \end{array}
    \right).
\end{equation}
This and equation \eqref{B} give the representation of level shift
operator as $\Lambda_0=B+A$.

\medskip

{\bf The level shift operators $\Lambda_{\pm\Delta}$.\ } In the
basis $\{\varphi_{12},\varphi_{13}\}$ of ${\rm Ran}P_\Delta$, we
have $\Lambda_\Delta=B+A$ with
 $$
B=\left(
   \begin{array}{cc}
     \Delta-\frac{\sigma}{2} & 0 \\
    0 & \Delta+\frac{\sigma}{2} \\
   \end{array}
 \right)
$$
and
 \begin{equation*}
 \begin{split}
 A\varphi_{12}=&-\lambda^2(2\langle \varphi(L_R+\i0^+)^{-1}\varphi\rangle-\overline{\langle \varphi(L_R+\i0^+)^{-1}\varphi\rangle} )
 \varphi_{12}+\lambda^2 \overline{\langle \varphi(L_R+\i0^+)^{-1}\varphi\rangle}\varphi_{13}\\
 A\varphi_{13}=&-\lambda^2(2\langle \varphi(L_R+\i0^+)^{-1}\varphi\rangle-\overline{\langle \varphi(L_R+\i0^+)^{-1}\varphi\rangle} )
 \varphi_{13}+\lambda^2 \overline{\langle \varphi(L_R+\i0^+)^{-1}\varphi\rangle}\varphi_{12}.
 \end{split}
 \end{equation*}
 So

  $$\Lambda_\Delta=\Delta-2\lambda^2\eta+\lambda^2\overline{\eta}\left(
                     \begin{array}{cc}
                      1 & 1 \\
                       1 & 1 \\
                     \end{array}
                   \right)+\sigma\left(
                                     \begin{array}{cc}
                                       -\frac12 & 0 \\
                                       0 & \frac12\\
                                     \end{array}
                                   \right),
 $$
 where $\eta=\langle \varphi(L_R+\i0^+)^{-1}\varphi\rangle$  and $\overline{\eta}$ is the complex conjugate of $\eta$.

Furthermore, one can easily see that
$\Lambda_{-\Delta}=-J\Lambda_\Delta J$ ($J$ is an anti-unitary
operator).

\subsubsection{Proof of Proposition \ref{prop1}}

Since $0<\sigma\ll\lambda^2\ll \Delta$, the operators $\Lambda_e$ can be viewed as a
perturbation of matrices $A$ by the small operators $B$. Analytic perturbation theory then gives the eigenvalues of $\Lambda_e$. For $\Lambda_0$:
\begin{equation}
\begin{split}
\varepsilon_0^{(1)} =&0, \\
\varepsilon_0^{(2)}
=&\i\frac{(2+\e^{-\beta\Delta})\delta}{2(1+\e^{-\beta\Delta})(4(Re\xi_-)^2+\delta^2)}\frac{\sigma^2}{\lambda^2}
+O(\lambda^2(\frac{\sigma}{\lambda^2})^4),\\
\varepsilon_0^{(3)} =&2\i\delta(1+e^{\beta\Delta})\lambda^2+O(\frac{\sigma^2}{\lambda^2}), \\
\varepsilon_0^{(4)} =&\i{\delta\lambda^2}+{2\lambda^2 \vartheta}+O(\frac{\sigma^2}{\lambda^2}),  \\
\varepsilon_0^{(5)} =&\i{\delta\lambda^2}-{2\lambda^2 \vartheta}+O(\frac{\sigma^2}{\lambda^2}),\\
\end{split}
\end{equation}
For $\Lambda_\Delta$, we obtain by analytic perturbation theory:
 \begin{eqnarray*}
 \varepsilon_1^{(1)} &=& \Delta -\lambda^2 {\rm P.V.}\int_{{\mathbb
R}^3}\frac{|g(k)|^2}{|k|}d^3k+2\i\lambda^2\frac{\widetilde J(0)}{\beta} +O(\frac{\sigma^2}{\lambda^2})\\
\varepsilon_1^{(2)} &=& \Delta+4\i\lambda^2\frac{\widetilde
J(0)}{\beta}+ O(\frac{\sigma^2}{\lambda^2}).
\end{eqnarray*}
Finally, the eigenvalues of $\Lambda_{-\Delta}$ are obtained by changing $\Delta$ to $-\Delta$ in the last formulas. This completes the proof of Proposition \ref{prop1}.

\subsubsection{Resonance projections}
\label{subsubresproj}

By analytic perturbation theory we find the eigenprojections of $\Lambda_0$ associated to
$\varepsilon_0^{(s)}$,
\begin{equation}
\label{2.36}
\begin{split}
 P_0^{(1)}=&(2+\e^{-\beta\Delta})^{-1}\left|\left[
   \begin{array}{c}
     1 \\
     1\\
    1 \\
0\\
0\\
   \end{array}
 \right]\right\rangle
\left\langle\left[
   \begin{array}{c}
     \e^{-\beta\Delta} \\
     1\\
    1 \\
0\\
0\\
   \end{array}
 \right]\right|\\
 =&(2+\e^{-\beta\Delta})^{-1}|(\varphi_{11}+\varphi_{22}+\varphi_{33})\rangle\langle(e^{-\beta\Delta}\varphi_{11}+\varphi_{22}+\varphi_{33})|\\
P_0^{(2)}=&(2+3\e^{-\beta\Delta}+\e^{-2\beta\Delta})^{-1}\left|\left[
   \begin{array}{c}
     1 \\
     1\\
    -1-\e^{-\beta\Delta} \\
0\\
0\\
   \end{array}
 \right]\right\rangle
\left\langle\left[
   \begin{array}{c}
     \e^{-\beta\Delta} \\
     1\\
    -1-\e^{-\beta\Delta} \\
0\\
0\\
   \end{array}
 \right]\right|+O(\sigma)\\
 =&(2+3\e^{-\beta\Delta}+e^{-2\beta\Delta})^{-1}|\varphi_{11}-\frac{\e^{-\beta\Delta}}{2}(\varphi_{22}+\varphi_{33})+(\varphi_{23}+\varphi_{32})(1+\frac
 {\e^{-\beta\Delta}}{2})\rangle\\
  &\langle |\frac{\e^{-\beta\Delta}}{2}(2\varphi_{11}-\varphi_{22}-\varphi_{33})+(\varphi_{23}+\varphi_{32})(1+\frac{\e^{-\beta\Delta}}{2})| +O(\sigma)\\
  P_0^{(3)}=&\langle u_3^*,u_3\rangle^{-1}|u_3\rangle\langle u_3^*|\\
 =&(1+\e^{-\beta\Delta})^{-1}\left|\left[
   \begin{array}{c}
     1 \\
     -e^{-\beta\Delta}\\
   0 \\
0\\
0\\
   \end{array}
 \right]\right\rangle
\left\langle\left[
   \begin{array}{c}
     1 \\
     -1\\
    0 \\
0\\
0\\
   \end{array}
 \right]\right|+O(\sigma)
\end{split}
\end{equation}
and
\begin{equation}
\begin{split}
 P_0^{(4)}=&\langle u_4^*,u_4\rangle^{-1}|u_4\rangle\langle u_4^*|\\
 =&\frac{1}{2}\left|\left[
   \begin{array}{c}
     0 \\
     0\\
   0 \\
1\\
1\\
   \end{array}
 \right]\right\rangle
\left\langle\left[
   \begin{array}{c}
     0 \\
     0\\
    0 \\
1\\
1\\
   \end{array}
 \right]\right|+O(\sigma),\\
 P_0^{(5)}=&\langle u_5^*,u_5\rangle^{-1}|u_5\rangle\langle u_5^*|\\
 =&\frac{1}{2}\left|\left[
   \begin{array}{c}
     0 \\
     0\\
   0 \\
1\\
-1\\
   \end{array}
 \right]\right\rangle
\left\langle\left[
   \begin{array}{c}
     0 \\
     0\\
    0 \\
1\\
-1\\
   \end{array}
 \right]\right|+O(\sigma)\\
\end{split}
\end{equation}
where $u_i$ ($u_i^*$) are eigenvectors of $\Lambda_0$
($\Lambda_0^*$) corresponding to $\varepsilon_0^{(s)} $ ($\overline
{\varepsilon_0^{(s)}}$), $s=3,4,5$.

Similarly, the eigenprojections of $\Lambda_{\pm\Delta}$ associated to
$\varepsilon_{\pm1}^{(1,2)}$ are
\begin{eqnarray}
P_{+1}^{(1)}&=&\frac{1}{2}|\varphi_{12}-\varphi_{13}\rangle \langle\varphi_{12}-\varphi_{13}| +O(\sigma) \\
P_{+1}^{(2)}&=&\frac{1}{2}|\varphi_{12}+\varphi_{13}\rangle
\langle\varphi_{12}+\varphi_{13}|+O(\sigma) 
\\
P_{-1}^{(1)}&=&\frac{1}{2}|\varphi_{21}-\varphi_{31}\rangle\langle
\varphi_{21}-\varphi_{31}|+O(\sigma) \\
P_{-1}^{(2)}&=&\frac{1}{2}|\varphi_{21}+\varphi_{31}\rangle
\langle\varphi_{21}+\varphi_{31}|+O(\sigma) 
\end{eqnarray}

\subsubsection{Proof of Proposition \ref{prop2}}
\label{subsupproofprop2}

Our starting point is \eqref{-1}. We have
\begin{equation}
\label{-3}
{\rm Tr}\rho_\infty A=\chi_0^{(1)}(A)+\chi_0^{(2)}(A) = \scalprod{\psi_S}{B(P^{(1)}_0+P^{(2)}_0)(A\otimes\bbbone_S)\psi_{\rm ref}}.
\end{equation}
The projections are given above, in Section \ref{subsubresproj} (with $\sigma=0$). Consider the term with $P^{(1)}_0$,
\begin{eqnarray}
\lefteqn{
 \scalprod{\psi_S}{B P^{(1)}_0(A\otimes\bbbone_S)\psi_{\rm ref}}}\label{-2}\\
 &=& \frac{{\scalprod{\psi_S}{B(\varphi_{11}+\varphi_{22}+\varphi_{33})} \scalprod{(e^{-\beta\Delta}\varphi_{11}+\varphi_{22}+\varphi_{33})}{(A\otimes\bbbone_S)\psi_{\rm ref}}}}{2+\e^{-\beta\Delta}}.
\nonumber
\end{eqnarray}
Using that
$$
\varphi_{jk}=\sqrt{3} (|\varphi_j\rangle\langle\varphi_k|\otimes\bbbone_S) \psi_{\rm ref}
$$
(see \eqref{refstate}), that $B$ commutes with $|\varphi_j\rangle\langle\varphi_k|\otimes\bbbone_S$ (see before \eqref{Bop}) and that $B\psi_{\rm ref}=\psi_S$ (see \eqref{Bop}), we have $\scalprod{\psi_S}{B\varphi_{jk}} = \sqrt{3}\scalprod{\psi_S}{(|\varphi_j\rangle\langle\varphi_k|\otimes\bbbone_S)\psi_S}=\sqrt{3}[\rho_0]_{k,j}$, which is the matrix element of the initials system state $\rho_0$ in the energy basis (see \eqref{9.1}). In particular,
\begin{equation}
\label{-4}
\scalprod{\psi_S}{B(\varphi_{11}+\varphi_{22}+\varphi_{33})} = \sqrt{3}.
\end{equation}
Next, from the expression \eqref{refstate} we obtain
\begin{equation}
\label{-5}
\scalprod{(e^{-\beta\Delta}\varphi_{11}+\varphi_{22}+\varphi_{33})}{(A\otimes\bbbone_S)\psi_{\rm ref}} =\frac{\e^{-\beta\Delta}A_{11}+A_{22}+A_{33}}{\sqrt 3},
\end{equation}
where $A_{kl}=\scalprod{\varphi_k}{A\varphi_l}$. Combining \eqref{-2}, \eqref{-4} and \eqref{-5} we obtain
\begin{equation}
 \scalprod{\psi_S}{B P^{(1)}_0(A\otimes\bbbone_S)\psi_{\rm ref}}={\rm Tr}(\rho_{S,\beta,0}A).
\label{equilauto}
\end{equation}
This gives one contribution, $\rho_{S,\beta,0}$, to $\rho_\infty$ (the one coming from $P_0^{(1)}$ in \eqref{-3}). One deals with $P^{(2)}_0$ in the same way and finds the result \eqref{20}.

The speed of convergence in \eqref{30} is given by the smallest non-zero imaginary part of all the $\varepsilon_j^{(s)}$ ($\sigma=0)$, which, according to Proposition \ref{prop1}, is proportional to $\lambda^2$. This shows Proposition \ref{prop2}.

{\em Remark.\ } The formula \eqref{equilauto} says that the contribution to the dynamics \eqref{-1} coming from one of the terms with vanishing resonance energy ($\varepsilon_0^{(1)}=0$) and associated projection $P_0^{(1)}$  yields simply the equilibrium part. This  general fact, not specific to the particular model at hand, is encoded by the fact that the Gibbs state of the system is always in the kernel (null-space) of the level shift operator associated to the unperturbed energy zero,  $\Lambda_0$ (see also \eqref{10}). The latter fact is, in turn, implied by the very construction of the Liouville operator, for which the {\em coupled} system-reservoir equilibrium state is an eigenvector with eigenvalue zero. We refer to \cite{MSB,M} for further detail on this.

\subsubsection{Stationarity of $\rho_{S,\beta,0}$ and $\rho_\tau$ for $\sigma=0$}
\label{subsubstat}

 In the Liouville representation of the 3-level system,
the Gibbs state $\rho_{S,\beta,0}$ is given by the normalized vector
$$\Omega_{S,\beta,0}={Z_\beta}^{-1/2}(\e^{-\beta E_0/2}\varphi_{11}+\e^{-\beta E/2}(\varphi_{11}+\varphi_{22})),$$
where $Z_\beta=\e^{-\beta E_0/2}+2\e^{-\beta E/2}.$ It follows from
\eqref{thechi} and \eqref{thenewdyn} that (note that $X_{\sigma=0}=\bbbone$)
\begin{equation}\label{42}
\begin{split}
{\rm Tr}(T_t(\rho_{S,\beta,0}) |\varphi_m\rangle\langle\varphi_l|) =&
\sum_{j=-1,0,1}\sum_{s=1}^{s_j} \e^{\i t\varepsilon_j^{(s)}}
\chi_j^{(s)}(|\varphi_m\rangle\langle\varphi_l|)\\
=&\sum_{j=-1,0,1}\sum_{s=1}^{s_j} \e^{\i t\varepsilon_j^{(s)}}
\langle \Omega_{S,\beta,0}, BP_j^{(s)}(|\varphi_m\rangle\langle\varphi_l|\otimes\bbbone)\psi_{{\rm ref}}\rangle.\\
\end{split}
\end{equation}
The operator $B$ is $B=\bbbone_S\otimes b$ with
$$
b=\sqrt{\frac{3}{\e^{\beta E}Z_\beta }}\left(
      \begin{array}{ccc}
        \e^{-{\beta\Delta}/2}    & 0         & 0\\
        0                              & 1         & 0 \\
        0                              & 0          & 1\\
      \end{array}
    \right).
$$
Using the explicit expressions of the resonance projections given in Section \ref{subsubresproj}, is easy to see that
$\chi_j^{(s)}(|\varphi_m\rangle\langle\varphi_l|)=0$ for all $j$ and
$s$, except for $j=0$ and $s=1,2$. Therefore, \eqref{42} becomes
\begin{equation}\label{43}
\begin{split}
{\rm Tr}(T_t(\rho_{S,\beta,0}) |\varphi_m\rangle\langle\varphi_l|)
={\rm Tr}(\rho_{S,\beta,0} |\varphi_m\rangle\langle\varphi_l|).
\end{split}
\end{equation}
Since \eqref{43} holds for all $m,l$, we have
$T_t(\rho_{S,\beta,0})=\rho_{S,\beta,0}$.

Next we consider $\rho_\tau$. In the Liouville representation space, the state $\rho_\tau$ is
represented by the normalized vector $\tau\otimes\tau.$ The associated operator
$B=\bbbone_S\otimes b$ satisfying $B\psi_{\rm ref}=\tau\otimes\tau$ is determined by
$$
b\varphi_1=0,\ b\varphi_2=\varphi_2-\varphi_3,\ b\varphi_3=\varphi_3-\varphi_2.
$$
Just as for $\rho_{S,\beta,0}$, this information together with the formulas in Section \ref{subsubresproj} gives $T_t(\rho_\tau)=\rho_\tau$.

\subsubsection{Proof of Proposition \ref{prop4}}

The result follows directly from \eqref{thenewdyn}, \eqref{thenewfinstate} and Proposition \ref{prop1}.

\appendix

\section{Spectral density and correlation function}
\label{appendix}

The definition of the spectral density is
\begin{equation}
\label{defspecdensity}
J(\omega) = \sqrt{2\pi}\tanh(\beta\omega/2) \,\widehat{\cal C}(\omega), \quad \omega\ge 0,
\end{equation}
where $\widehat {\cal C}(w)$ is the Fourier transform,
$$
\widehat {\cal C}(w) = \frac{1}{\sqrt{2\pi}}\int_{-\infty}^\infty \e^{-\i wt}{\cal C}(t)\d t,\quad w\in{\mathbb R},
$$
of the {\em symmetrized correlation function}
\begin{equation}
\label{cete}
{\cal C}(t) = \frac12 \Big[\big\langle \e^{\i t H_R }\varphi(g)\e^{-\i t H_R }\varphi(g)\big\rangle_\beta + \big\langle \varphi(g)\e^{\i t H_R }\varphi(g)\e^{-\i t H_R }\big\rangle_\beta\Big].
\end{equation}
Here, $g$ is the form factor of the interaction (see \eqref{4}) and $\langle\cdot\rangle_\beta$ is the thermal average in the reservoir equilibrium state. Note that 
$$
{\cal C}(t)=\overline{\cal C}(t) = {\cal C}(-t)\qquad \mbox{and}\qquad \widehat{\cal C}(\omega)=\overline{\widehat{\cal C}}(\omega) = \widehat{\cal C}(-\omega).
$$
The definition \eqref{defspecdensity} is the same as in \cite{MBSa}. In the latter paper, it is shown (Section 4) that it coincides with the definition of the reservoir spectral density for the discretized modes given in \cite{Leggett}. A direct calculation of the correlation function and its Fourier transform, together with \eqref{defspecdensity} yields the expression \eqref{specdensity}. We point out that slightly different conventions are used in the literature; for instance the spectral density of \cite{MBR} is {\em twice} that used here. An explicit calculation gives
\begin{equation}
\label{a1}
{\cal C}(t) =\frac12 {\rm Re} \int_{\mathbb R} \e^{-\i ut} \frac{u^2}{|1-\e^{-\beta u}|} \int_{S^2} |g(|u|,\Sigma)|^2 \d\Sigma \ \d u.
\end{equation}

\medskip
\noindent
{\bf Reservoir correlation time $\tau_{\rm c}$.\ } Take 
\begin{equation}
\label{a0}
g(k)=A|k|^p\e^{-\frac12(|k|/\varkappa_0)^m},
\end{equation}
for some amplitude $A$ and $p=-1/2+n$, $n=0,1,2,\ldots$, $\varkappa_0>0$ and $m=1,2$ (see before \eqref{7.1}). Then we obtain from \eqref{a1} that 
\begin{equation}
\label{a2}
{\cal C}(t) =2\pi A^2 {\rm Re}\int_{\mathbb R} \e^{-\i ut}\frac{|u|^{2(1+p)}}{|1-\e^{-\beta u}|} \e^{-(|u|/\varkappa_0)^m} \d u.
\end{equation}
We wish to obtain the decay properties of ${\cal C}(t)$, as $t$ becomes large. Consider $p=-1/2$, which is an important case since then $\widetilde J(0)>0$ (see \eqref{xizero} and also Proposition \ref{prop4}). For definiteness, let $m=1$ in \eqref{a0}. Then
\begin{equation}
\label{a10}
{\cal C}(t) =2\pi A^2 \beta^{-2} {\rm Re}\int_{\mathbb R} \e^{-\i ut/\beta}\left| \frac{u}{1-\e^{-u}}\right| \e^{-|u|/(\beta \varkappa_0)} \d u.
\end{equation}
Splitting the integral in \eqref{a10} over positive and negative values of $u$ and using that
$$
\left| \frac{u}{1-\e^{-u}}\right| =
\left\{
\begin{array}{ll}
u + u\frac{\e^{-u}}{1-\e^{-u}}, & u\ge 0\\
\frac{u}{1-\e^{-u}}, & u<0
\end{array}
\right.
$$
we arrive at
\begin{equation}
\label{a10.00}
{\cal C}(t) =2\pi A^2 \beta^{-2} [T_1(t)+T_2(t)].
\end{equation}
Here,
\begin{equation}
\label{a10.01}
T_1(t) = {\rm Re}\int_0^\infty \e^{-\i ut/\beta} u\e^{-u/(\beta\varkappa_0)}\d u = (\beta\varkappa_0)^2\frac{1-(t\varkappa_0)^2}{(1+(t\varkappa_0)^2)^2}
\end{equation}
and 
\begin{equation}
\label{a10.02}
T_2(t) = 2{\rm Re}\int_0^\infty \e^{-\i ut/\beta} \frac{u}{1-\e^{-u}}\e^{-u} \e^{-u/(\beta\varkappa_0)}\d u.
\end{equation}
In the integral \eqref{a10.02}, only values of $0\le u\le 1$ contribute essentially, due to the decay of the factor $\e^{-u}$. Also, for these $u$, we can replace $\frac{u}{1-\e^{-u}}\approx 1$. Then we readily obtain
\begin{equation}
\label{a10.03}
T_2(t) = 2 \beta\varkappa_0 \frac{\beta\varkappa_0+1}{(\beta\varkappa_0+1)^2 +(t\varkappa_0)^2}.
\end{equation}
{}From \eqref{a10.00}, \eqref{a10.01} and \eqref{a10.03}, we see that the reservoir correlation time is 
\begin{equation}
\label{a13}
\tau_{\rm c} = \frac{1}{\varkappa_0}.
\end{equation}

{\em Remark.\ } An equivalent (discrete mode) expression for the correlation function \eqref{cete} is 
\begin{equation}
\label{a0001}
{\cal C}(t) = \tfrac12\sum_{\vec{k}}|g_{\vec{k}}|^2\cos(\omega_{\vec{k}} t)[2n_{\vec{k}} +1],
\end{equation}
where $n_{\vec{k}}$ is the average of the number of reservoir modes with given discretized wave vector ${\vec k}\in{\mathbb R}^3$. For a thermal reservoir, $n_{\vec{k}}=\frac{1}{\e^{\beta \omega_{\vec{k}}}-1}$. Consider the dispersion relation $\omega_{\vec{k}}=k\equiv|\vec{k}|$. Then $2n_{\vec{k}}+1=\coth(\beta k/2)$. Due to the quantum vacuum energy, represented by the term $+1$ in \eqref{a0001}, the thermal factor $2n_{\vec{k}}+1$ does not cut off high frequencies, as $2n_{\vec{k}}+1\approx 1$ as $k\rightarrow\infty$. This is why our reservoir correlation decay time is $1/\varkappa_0$, given by the inverse of the cutoff frequency we have introduced by hand in the form factor \eqref{a0}. For a classical reservoir, instead of the termal factor $2n_{\vec{k}}+1$ in \eqref{a0001} we would only have $2n_{\vec{k}}\sim 2e^{-\beta \omega_k}$. In that situation, one can check that the decay time of ${\cal C}(t)$ is different for small and large temperatures. Namely, the decay time is $1/\varkappa_0$ if $\beta\varkappa_0<\!\!<1$ but for low temperatures, when $\beta\varkappa_0>\!\!>1$, the decay time is $1/\beta$, since the available modes are cut off by the thermal distribution.

\bigskip
\noindent
{\bf System decay times in terms of reservoir correlation time.} With the choice $p=-1/2$, $m=1$, we obtain from \eqref{specdensity}  and \eqref{a13} that
\begin{equation}
\label{a5}
J(\omega)=2\pi^2A^2 \omega \e^{-\omega/\varkappa_0} = 2\pi^2A^2 \omega \e^{-\omega\tau_{\rm c}}.
\end{equation}
The resevoir spectral density $J(\omega)$ hence depends on the reservoir correlation time $\tau_{\rm c}$. Consequently, so do the characteristic decay rates of the open system, see \eqref{gammadeg} and \eqref{gammand}. The definition of $\widetilde J(0)$, \eqref{xizero}, together with \eqref{a5} shows that 
\begin{equation}
\label{a14}
\widetilde J(0) = 2\pi^2 A^2
\end{equation}
does not depend on $\tau_{\rm c}$, while 
$$
J(\Delta)=2\pi^2A^2 \Delta \e^{-\Delta\tau_{\rm c}}.
$$
Note that the decay rate for the non-degenerate system, \eqref{gammand}, has a dependence on $\tau_{\rm c}$ not only via the spectral density $J(\Delta)$, but for small temperatures also via the constant $\vartheta$.

\bigskip
\noindent
{\bf Acknowledgments.\ } M.M. and H.S. have been supported by NSERC through a Discovery Grant. M.M. is grateful for the hospitality and financial support of the Center for Nonlinear Studies of LANL, where part of this work was carried out. We are grateful to two anonymous referees for a very careful reading of our paper and for giving us valuable comments.

\end{document}